\documentclass[titlepage,draft,11pt]{article} 
\usepackage{amsfonts,latexsym} 
\usepackage{pstricks,pst-plot}
\usepackage{amsmath} 
\usepackage{amsthm}
\usepackage[a4paper]{geometry}

\date{}

\newtheorem{thm}{Theorem}[section]

\newtheorem{rmk}[thm]{Remark}

\title{ Recovering the  relativistic kinetic energy.}

\author{Alain Haraux\vspace{1ex}\\ 
{\normalsize Sorbonne Universit\'e, Universit\'e Paris-Diderot SPC, CNRS, INRIA}, \\
{\normalsize Laboratoire Jacques-Louis Lions,  LJLL, F-75005,
Paris, France.}\\ 
{\normalsize e-mail: \texttt{alain.haraux@sorbonne-universite.fr}}}

\begin{document}
\maketitle
\begin{abstract}
We recover the relativistic kinetic energy as the result of the work of a force. \\

\vspace{1cm} 

 \end{abstract}

%%%%%%%%%%%%%%%%%%%%%
%                   %
%   Inizio lavoro   %
%                   %
%%%%%%%%%%%%%%%%%%%%%
 
\section{Introduction}
In special relativity, it is assumed that the mass of a moving massive object evaluated in a given spatial reference frame ${\cal R}$ is given by the formula  \begin{equation}\label{dil}  m = m(v) =  \frac{m_0} {\sqrt{ 1-\frac{v^2}{c^2}}}\end{equation} where $m_0$ is the mass of the same object at rest in ${\cal R}$  and $v, c$ are respectively the scalar velocity of the object measured in ${\cal R}$  and the scalar velocity of light. This formula refers to the so-called mass dilation property.  A naive approach to  the kinetic energy of that object  by imitating  the Newtonian framework would give  the formula 
\begin{equation*}\label{KE}KE = \frac{1}{2} m(v)v^2= \frac{1}{2}\frac{m_0 v^2} {\sqrt{ 1-\frac{v^2}{c^2}}}, \end{equation*} However in the framework of special relativity (SR), the formula which is used is  
\begin{equation}\label{KE*}KE_{rel} =\left[\frac{1} {\sqrt{ 1-\frac{v^2}{c^2}}}-1\right]m_0 c^2 = (\gamma-1)m_0 c^2 \end{equation} where $\gamma = \frac{1}{\sqrt{ 1-\frac{v^2}{c^2}}}$ is the Lorentz factor, which gives a kinetic energy about twice as large as $v$ approaches $c$. This is usually justified by taking the difference between the total energy at speed $v$ and the total energy at rest through the celebrated formula $ E= mc^2$.  \\

\noindent
Actually, as shown in Section 2, one can recover the relativistic kinetic energy of any massive object from the total work of forces used to pass from rest to velocity $v$.   As a complement, in Section 3 we compute the trajectory of a mass subject to an exterior force $F$ in the relativistic framework. 

\section{A proof relying on mechanical work.}   
 
Let us consider a massive object initially at rest (at time $t= 0$ in  ${\cal R}$  with mass $m_0$, and with final velocity $v= v(t)$ obtained by the action of a force $$F = F(s), \quad 0\le s\le t.$$ We start from Newton's second law in the form \begin{equation}\label{N2} F(s) = p'(s) = \frac{d}{ds}\left[m(v(s))v(s)\right] \end{equation} and, by the conservation of energy principle, we infer that the kinetic energy ${\cal E}$ acquired by the massive object is equal to the total mechanical work of the force $F$, which provides the formula 
\begin{equation}\label{E} {\cal E} = {\cal E}(t) = \int_0^t (F (s), dx(s)) = \int_0^t  (\frac{d}{ds}\left[m(v(s))v(s)\right],  v(s))ds \end{equation} Integrating by parts, we obtain 
$$ {\cal E}(t) = m(v(t))||v(t)||^2 - \int_0^t m(v(s))  (v(s), v'(s)) ds  $$ On the other hand we have $$ (v(s), v'(s) = \frac{1}{2} \frac{d}{ds} (||v(s)||^2)$$ so that 

$$ \int_0^t m(v(s))  (v(s), v'(s)) ds = m_0 \int_0^t \frac {\frac{d}{ds} (||v(s)||^2)} {2\sqrt{ 1-\frac{||v(s)||^2}{c^2}}} ds $$ $$ = -m_0c^2 \int_0^t \frac{d}{ds} \left[ \sqrt{ 1-\frac{||v(s)||^2}{c^2}}\right] ds = m_0c^2 \left(1-  \sqrt{ 1-\frac{||v(t)||^2}{c^2}}\right)$$
 We end up with $$ {\cal E}(t) = m_0 \frac{||v(t)||^2}{ \sqrt{ 1-\frac{||v(t)||^2}{c^2}}} + m_0c^2 \sqrt{ 1-\frac{||v(t)||^2}{c^2}}- m_0c^2$$ Introducing  $v: = ||v(t)||$,  the {\it scalar velocity } at time $t$, it turns out that $$ \frac{ v^2} {\sqrt{ 1-\frac{v^2}{c^2}}} + c^2 {\sqrt{ 1-\frac{v^2}{c^2}}} = \frac{ v^2 + c^2 (1-\frac{v^2}{c^2})} {{\sqrt{ 1-\frac{v^2}{c^2}}} } = \frac{ c^2 } {{\sqrt{ 1-\frac{v^2}{c^2}}} }$$ Hence finally 
\begin{equation}\label{E2} {\cal E} = {\cal E}(t) = \left[\frac{1} {\sqrt{ 1-\frac{v^2}{c^2}}}-1\right]m_0 c^2\end{equation} 
 We recover \eqref{KE*}.
\noindent We now make a few remarks. 
\begin{rmk} As expected, the result  depends neither  on the path followed by the massive object, nor on the evolution of $v(s)$ on the time interval $[0, t]$, it only depends on the mass $m_0$  at rest and the final scalar velocity $v$. \end{rmk} 
\begin{rmk} If $v<<c$, we have $ \displaystyle {\cal E} (m_0, v)\sim \frac{m_0v^2}{2}$ as in the case of classical Newtonian mechanics. \end{rmk} 
\begin{rmk}The author was induced to compute the relativistic kinetic energy when trying to understand the origin of inertia. That problem seems to be one of the most important of fundamental physics, together with the origin of the gravitational force, cf. about this the references \cite{Mach,  Sci, VanF} . \end{rmk}

\section{The evolution of velocity in the forced case}  Here we compute the evolution of the velocity $v$ for a point mass subject to a force $F$ with initial velocity $0$.  Equation \eqref{N2} gives 
$$ m(v(t)) v(t) = \int_0^t F(s) ds := G(t) $$ Taking the square of the norm on both sides of the equation yields, keeping the simplified notation $v = ||v(t)||$, the simple formula 
\begin{equation}\label{v2}  v^2 = \frac{ c^2} {1+ \frac{m_0c^2}{||G(t)||^2} }\end{equation} Which gives the evolution of the scalar velocity \begin{equation}\label{v}  ||v(t)|| = \frac{ c} {\sqrt{1+ \frac{m_0 c^2}{||G(t)||^2} }}\end{equation} More precisely we have \begin{equation}\label{v(t)}  v(t) = \frac{  c\, G(t) } {\sqrt{||G(t)||^2+ {m_0 c^2}}}\end{equation} For instance if $F$ is a constant vector, $v$ tends to $c$ as t tends to infinity, but $v$ never equals the velocity of light.

\begin{rmk} When F(t) = F is constant, the difference $ |v(t)-c\frac{F}{||F||}| $ tends to $0$ as $\frac{1}{t^2}$. \end{rmk}
\begin{rmk} If we make $c\to\infty$ in \eqref{v(t)} , we recover the formula \begin{equation*} v(t) = \frac{ G(t) } {m_0 }.\end{equation*} \end{rmk}

\bigskip \noindent {\bf Acknowledgement} The author is indebted to Anne Adamczewsky for pointing out a basic mistake in the first version of this preprint.

\end{document}